\documentclass[conference]{IEEEtran}
\IEEEoverridecommandlockouts

\usepackage{url}
\usepackage{cite}
\usepackage{amsmath,amssymb,amsfonts}
\usepackage{algorithmic}
\usepackage{graphicx}
\usepackage{textcomp}
\usepackage{xcolor}
\usepackage{hyperref}
\usepackage{todonotes}
\setuptodonotes{inline}

\def\BibTeX{{\rm B\kern-.05em{\sc i\kern-.025em b}\kern-.08em
    T\kern-.1667em\lower.7ex\hbox{E}\kern-.125emX}}

\begin{document}

\title{Towards Evaluation Guidelines for Empirical Studies involving LLMs

\thanks{The work of Davide Falessi has been partially supported by the 2022JJ3PA5 - PRIN2022 project. This work also was partially supported by the German Federal Ministry of Education and Research in the project MEKI (21IVP016F).}
}

\author{
    \IEEEauthorblockN{Stefan Wagner\IEEEauthorrefmark{1}, Marvin Muñoz Barón\IEEEauthorrefmark{1}, Davide Falessi\IEEEauthorrefmark{2}, Sebastian Baltes\IEEEauthorrefmark{3}}
    \IEEEauthorblockA{\IEEEauthorrefmark{1}TUM School of Computation, Information and Technology \\
    Technical University of Munich, Heilbronn, Germany\\
    \{stefan.wagner, marvin.munoz-baron\}@tum.de}
    \IEEEauthorblockA{\IEEEauthorrefmark{2}University of Rome ``Tor Vergata'', Rome, Italy\\
    falessi@ing.uniroma2.it}
    \IEEEauthorblockA{\IEEEauthorrefmark{3}University of Bayreuth, Bayreuth, Germany\\
    sebastian.baltes@uni-bayreuth.de} 
}


\maketitle

\begin{abstract}
In the short period since the release of ChatGPT, large language models (LLMs) have changed the software engineering research landscape.
While there are numerous opportunities to use LLMs for supporting research or software engineering tasks, solid science needs rigorous empirical evaluations.
However, so far, there are no specific guidelines for conducting and assessing studies involving LLMs in software engineering research.
Our focus is on empirical studies that either use LLMs as part of the research process or studies that evaluate existing or new tools that are based on LLMs.
This paper contributes the first set of holistic guidelines for such studies.
Our goal is to start a discussion in the software engineering research community to reach a common understanding of our standards for high-quality empirical studies involving LLMs.
\end{abstract}


\section{Introduction}

While artificial intelligence (AI) has been used in software engineering (SE) for a long time, success used to be limited \cite{DBLP:journals/tosem/Martinez-Fernandez22}.
Recently, the rise of large language models (LLMs) has opened new avenues for the application of AI in software engineering~\cite{fan2023large,10.1145/3695988}. These models offer many possible use cases, ranging from code generation and bug detection to requirements analysis and software maintenance. For instance, LLM-based tools were able to generate logging statements~\cite{li2024exploring}, generate test cases~\cite{santos2024we}, and support education~\cite{10663055}.

As a result, we are starting to see an increasing number of evaluation studies either using LLMs as part of the research process~\cite{DBLP:journals/ase/BanoHZT24} or as part of tools that automate or improve software engineering tasks. These studies explore the effectiveness, performance, and robustness of LLMs in different contexts, such as improving code quality, reducing development time, or supporting software documentation. However, it is often unclear how valid and reproducible results can be achieved with empirical studies involving LLMs -- or what effect their usage has on the validity of empirical results. This uncertainty poses significant challenges for researchers aiming to draw reliable conclusions from empirical studies.

One of the primary risks in creating unreproducible results stems from the variability in LLM performance due to differences in training data, model architecture, evaluation metrics, and the inherent non-determinism of those models. For example, slight changes in the training data or the hyperparameters can lead to significantly different outcomes, making it difficult to reproduce studies.
Also, the lack of standardized benchmarks and evaluation protocols further complicates the reproducibility. These issues highlight the need for clear guidelines and best practices to ensure that empirical studies with LLMs yield valid and reproducible results.

There has been extensive work developing guidelines for conducting and reporting specific types of empirical studies such as controlled experiments \cite{DBLP:books/sp/WohlinRHORW24, DBLP:books/sp/08/JedlitschkaCP08} or their replications \cite{DBLP:journals/tse/SantosVOJ21}. We believe that LLMs have specific intrinsic characteristics that require specific guidelines for researchers to achieve an acceptable level of reproducibility.
For example, even if we know the specific version of an LLM used for an empirical study, the reported performance for the studied tasks can change over time, especially for commercial models that evolve beyond version identifiers~\cite{DBLP:journals/corr/abs-2307-09009}.
Moreover, commercial providers do not guarantee the availability of old model versions indefinitely.
Besides versions, LLMs' performance widely varies depending on configured parameters such as temperature. Therefore, not reporting the parameter settings impacts the reproducibility of the research. 

Even for ``open'' models such as Llama, we do not know how they were fine-tuned for specific tasks and what the exact training data was~\cite{Gibney2024}.
For example, when evaluating LLMs' performance for certain programming tasks, it would be relevant to know whether the solution to a certain problem was part of the training data or not.




Therefore, with this paper, we provide two key contributions: (1) a classification of different types of empirical studies involving LLMs in software engineering research and (2) preliminary guidelines on how to achieve valid and reproducible results in such studies.
The most recent version of the study types and guidelines is available online.\footnote{\url{https://llm-guidelines.org/}}



\section{Related Work}

There are several established guidelines for empirical studies in software engineering, e.g., for experiments~\cite{DBLP:books/sp/WohlinRHORW24,DBLP:books/sp/08/JedlitschkaCP08}. While these guidelines continue to be useful, they were developed before the rise of LLMs. Therefore, this paper is a starting point for extending the existing set of guidelines.
To the best of our knowledge, the only similar work is a paper by Sallou, Durieux, and Panichella~\cite{sallou2024breaking}, in which they also call for a broader discussion in the community.
They mostly focus on a discussion of threats to validity, proposing guidelines that partly overlap with ours.
However, they do not structure their guidelines according to different types of studies.
We are convinced that the diversity of studies involving LLMs requires a differentiation between study types.
Our taxonomy presented in Section~\ref{sec:study-types} is a first step in that direction.





\section{Types of Studies}
\label{sec:study-types}




The development of empirical guidelines for studies involving LLMs in software engineering is crucial for ensuring the validity and reproducibility of results. However, these guidelines must be tailored for different study types as they may pose unique challenges. Therefore, understanding the classification of these studies is essential for developing appropriate guidelines.
We envision that a mature set of guidelines provides specific guidance for each of these study types, addressing their individual methodological idiosyncrasies.

\subsection{LLMs as Tools for Researchers in Empirical Studies}

LLMs can be leveraged as powerful tools to assist researchers conducting empirical studies. They can automate various tasks such as data collection, preprocessing, and analysis. For example, LLMs can extract relevant information from large datasets, generate summaries of research papers, and even assist in writing literature reviews. This can significantly reduce the time and effort required by researchers, allowing them to focus on more complex aspects of their studies.

\subsubsection{LLMs as Annotators}
LLMs can serve as annotators by automatically labeling artifacts with corresponding categories for data analysis.
For example, in a study analyzing code changes in version control systems, researchers may need to categorize each individual change.
For that, they may use LLMs to analyze commit messages and categorize them into predefined labels such as bug fixes, feature additions, or refactorings.
This automation can improve the efficiency of the annotation process, which is often a labor-intensive and error-prone task when done manually.
Moreover, in qualitative data analysis, manually annotating or coding text passages is also an often time-consuming manual process.
LLMs can be used to augment human annotations, provide suggestions for new codes, or even automate the entire process.
In such tasks, LLMs have the potential to improve the accuracy and efficiency of automated labeling processes~\cite{wan2024tnt}, making them valuable tools for empirical research in software engineering.
Hybrid human-LLM annotation approaches may further increase accuracy and allow for the correction of incorrectly applied labels~\cite{wang2024human}.

\subsubsection{LLMs as Judges}

In empirical studies, LLMs can act as judges to evaluate the quality of software artifacts such as code, documentation, and design patterns. 
For instance, LLMs can be trained to assess code readability, adherence to coding standards, and the quality of comments. 
By providing rather objective and consistent evaluations, LLMs could help mitigate certain biases and part of the variability that human judges might introduce. 
This could lead to more reliable and reproducible results in empirical studies.
However, when relying on the judgment of LLMs, researchers have to make sure to build a reliable process for generating ratings that considers the non-deterministic nature of LLMs and report the intricacies of that process transparently.

\subsubsection{LLMs as Subjects}

LLMs can be used as subjects in empirical studies to simulate human behavior and interactions. For example, researchers can use LLMs to generate responses in user studies, simulate developer interactions in collaborative coding environments, or model user feedback in software usability studies. This approach can provide valuable insights while reducing the need to recruit human participants, which can be time-consuming and costly. Additionally, using LLMs as subjects allows for controlled experiments with consistent and repeatable conditions.
However, when using LLMs as study subjects, it is important that researchers are aware of their inherent biases~\cite{Crowell2023} and limitations~\cite{DBLP:journals/ais/HardingDLL24}.

\subsection{LLMs for New Tools Supporting Software Engineers}

LLMs are being integrated into new tools designed to support software engineers in their daily tasks. These tools can include intelligent code editors that provide real-time code suggestions, automated documentation generators, and advanced debugging assistants. Empirical studies can evaluate the effectiveness of these tools in improving productivity, code quality, and developer satisfaction. By assessing the impact of LLM-powered tools, researchers can identify best practices and areas for further improvement.
For example, Choudhuri et al.~\cite{choudhuri2024far} conducted a student experiment in which they measured the impact of ChatGPT on the correctness and completion time for programming tasks.

\subsection{Studying LLM Usage}

Empirical studies can also focus on understanding how software engineers use LLMs in their workflows. This involves investigating the adoption, usage patterns, and perceived benefits and challenges of LLM-based tools. Surveys, interviews, and observational studies can provide insights into how LLMs are integrated into development processes, how they influence decision-making, and what factors affect their acceptance and effectiveness. Such studies can inform the design of more user-friendly and effective LLM-based tools.
For example, Khojah et al.~\cite{khojah2024beyond} investigated the use of ChatGPT by professional software engineers in a week-long observational study.

\subsection{Benchmarking LLMs for SE Tasks}
Another typical type of study focuses on benchmarking LLM output quality on large datasets.
In software engineering, this may include the evaluation of LLMs' ability to produce accurate and robust outputs for input data from real-world projects or synthetically created SE datasets.
In studies with generative models, the LLM output is often compared against a ground truth dataset using similarity metrics such as ROUGE, BLEU, or METEOR~\cite{10.1145/3695988}.
Moreover, the evaluation may be augmented by using task-specific or artifact-specific measures. 
Such measures may include code quality or performance metrics for code generation tasks or readability metrics for natural language SE artifacts (e.g., requirements documents). 
In this context, reference datasets such as HumanEval~\cite{chen2021evaluating} play an important role in establishing standardized evaluations. 
However, benchmark contamination~\cite{ahuja2024contaminationreportmultilingualbenchmarks} has recently been identified as an issue.
The careful creation of samples and corresponding input prompts is particularly important, as correlations between prompts may bias benchmark results~\cite{siska2024examining}.

\section{Preliminary Guidelines}

While providing a comprehensive set of guidelines is beyond the scope of this position paper, we report a first set of guidelines based on a discussion session with other empiricism experts at the 2024 International Software Engineering Research Network (ISERN) meeting.\footnote{\url{https://isern.fraunhofer.de}} 
This paper is meant as a starting point for further discussions in the community with the aim of developing a common understanding of how we should conduct and report empirical studies involving LLMs.

\subsection{Declare LLM Usage and Role}

When conducting any kind of empirical study involving LLMs, it is essential to clearly declare that an LLM was used. This includes specifying the purpose of using the LLM in the study, the tasks it was applied to, and the expected outcomes. Transparency in the usage of LLMs helps in understanding the context and scope of the study, facilitating better interpretation and comparison of results.
Beyond this declaration, we recommend that the authors be explicit about the LLM's exact role.
Oftentimes, there is a complex layer around the LLM that preprocesses data, prepares prompts, or filters user requests.
One example is ChatGPT, which can, among others, use the GPT-4o model.
GitHub Copilot uses the same model as well, and researchers can build their own tools utilizing GPT-4o directly (e.g., via the OpenAI API).
The infrastructure around the bare model can significantly contribute to the performance of a model in a certain task.
Therefore, it is crucial that researchers clearly describe what the LLM contributes to the tool or method presented in a research paper.


\subsection{Report Model Version and Date}

It is also crucial for all types of studies to document the specific version of the LLM used in the study, along with the date when the experiments were conducted. LLMs are frequently updated, and different versions may produce varying results. 
By providing this information, researchers enable others to reproduce the study under the same conditions. Different model providers have varying degrees of information. For example, OpenAI provides a model version and a system fingerprint describing the backend configuration that can also influence the output. Therefore, stating ``We used gpt-4o-2024-08-0, system fingerprint fp\_6b68a8204b'' provides clarity on the exact model and runtime environment. 
However, the main purpose of the system fingerprint is detecting changes and going back to a previous system fingerprint is impossible. 

\subsection{Report Model Configuration}

Detailed documentation of the configuration and parameters used during any study is necessary for reproducibility. This includes settings such as the temperature that controls randomness, the maximum token length, and any other relevant parameters such as the consideration of historical context.
Additionally, a thorough description of the hosting environment of the LLM or LLM-based tool should be provided, especially in studies focusing on performance or any time-sensitive measurement.
For instance, researchers might report that ``the model was integrated via the Azure OpenAI Service, and configured with a temperature of 0.7, top\_p set to 0.8, and a maximum token length of 512,'' providing a clear overview of the experimental setup.
Using seed values does further increase reproducibility, but does not completely mitigate the issue of non-determinism.\footnote{\href{https://cookbook.openai.com/examples/reproducible_outputs_with_the_seed_parameter}{Open AI Cookbook: The new seed parameter}}

\subsection{Report Prompts and their Development}

Reporting the exact prompts used in the study is essential for transparency and reproducibility.
Prompts can significantly influence the output of LLMs~\cite{Liu:2024:Refining}, and sharing them allows other researchers to understand and reproduce the conditions of the study.
For example, including the specific questions or tasks given to the LLM helps assess the validity of the results and compare them with other studies.
This is an example where different types of studies require different information.
When studying LLM usage, the researchers ideally collect and publish the prompts written by the users (if confidentiality allows).
Otherwise, summaries and examples can be provided.
Prompts also need to be reported when LLMs are integrated into new tools, especially if study participants were able to formulate (parts of) the prompts.
For all other types of studies, researchers should discuss how they arrived at their final set of prompts.
If a systematic approach was used, this process should be described in detail.

\subsection{Use an Open LLM as a Baseline}
To ensure the reproducibility of results, we recommended findings be reported with an open LLM as a baseline.
This applies both when using LLMs as tools for supporting researchers in empirical studies and when benchmarking LLMs for SE tasks.
In case LLMs are integrated into new tools, this is also preferable if the architecture of the tool allows it.
If the effort of changing models is too high, researchers should at least report an initial benchmarking with open models, which enables more objective comparisons.
Open LLMs can either be hosted via cloud platforms such as \emph{Hugging Face} or used locally via tools such as \emph{ollama} or \emph{LM Studio}.
A replication package for papers using LLMs should include clear instructions that allow other researchers to reproduce the findings using open models.
This practice enhances the credibility of the study and allows for independent verification of the results. 
Researchers could, e.g., mention that ``results were compared with those obtained using Meta’s Code LLAMA, available on the Hugging Face platform'' and point to a replication package.

We are aware that the definition of an ``open'' model is actively being discussed, and many open models are essentially only ``open weight''~\cite{Gibney2024}.
We consider the \emph{Open Source AI Definition} proposed by the \emph{Open Source Initiative} (OSI) to be a first step towards defining true open-source models~\cite{OSIAI2024}.

\subsection{Use Human Validation for LLM Outputs}

Especially in studies where LLMs are used to support researchers, human validation should always be employed.
While LLMs can automate many tasks, it is important to validate their outputs with human annotations, at least partially. For natural language processing tasks, a large-scale study has shown that LLMs have too large a variation in their results to be reliably used as a substitution for human judges~\cite{bavaresco2024llms}. Human validation helps ensure the accuracy and reliability of the results, as LLMs may sometimes produce incorrect or biased outputs. Incorporating human judgment in the evaluation process adds a layer of quality control and increases the trustworthiness of the study’s findings, especially when explicitly reporting inter-rater reliability metrics. For instance, ``A subset of 20~\% of the LLM-generated annotations were reviewed and validated by experienced software engineers to ensure accuracy. An inter-rater reliability of 90~\% was reached.''
For studies using LLMs as annotators, the proposed process by Ahmed et al.~\cite{ahmed2024can}, which includes an initial few-shot learning and, given good results, the replacement of \emph{one} human annotator by an LLM, might be a way forward.



\section{Conclusions}

In this paper, we outlined preliminary guidelines for researchers reporting on empirical studies involving Large Language Models (LLMs) in software engineering research.
While researchers can already use these guidelines to improve the reproducibility of their studies and the reporting in their papers, we see a demand for more tailored guidelines focusing on the different study types we identified in Section~\ref{sec:study-types} and also for extending the guidelines to include missing aspects such as ethical implications of using LLMs in research~\cite{ungless2024ethics}.
One aspect to focus on could be the particular types of threats to validity that arise from using LLMs in the context of different study types, building on related work~\cite{sallou2024breaking}.
Another direction could be to conduct a critical review of published studies involving LLMs, assessing how many papers already adhere to the guidelines we suggest.

\bibliographystyle{ieeetr}
\bibliography{literature}

\end{document}